\input epsf.sty
\newdimen\psfigsize
\def\psfigure#1 #2 #3 #4 #5{
    \begin{figure}[tbh]
      \vbox{
        \null\vskip-0.1in\hskip#2
        \epsfxsize=#1
        \epsfbox{#4}
        \vskip -0.1in
        \caption {#5 \label{#3}}
        \vskip 0.0truein plus0.1truein
      }
    \end{figure}
}
\def\nab#1{{\nabla_{#1}}}
\def\nabstar#1{\nabla\kern-0.5pt\smash{\raise 4.5pt\hbox{$\ast$}}
               \kern-4.5pt_{#1}}

\def\drvstar#1{\partial\kern-0.5pt\smash{\raise 4.5pt\hbox{$\ast$}}
               \kern-5.0pt_{#1}}

\def\newline{\relax\ifhmode\null\hfil\break\else\nonhmodeerr@\newline\fi}
\def\frac#1#2{{#1\over#2}}
\def\text#1{{\hbox{\rm #1}}}

\newcommand{\beq}{\begin{equation}}
\newcommand{\eeq}{\end{equation}}
\newcommand{\bea}{\begin{eqnarray}}
\newcommand{\eea}{\end{eqnarray}}
\def\Id{ \mbox{1\hspace{-1.2mm}I} }
\def\BE{\begin{equation}}
\def\EE{\end{equation}}
\def\BA{\begin{eqnarray}}
\def\EA{\end{eqnarray}}
\def\BAN{\begin{eqnarray*}}
\def\EAN{\end{eqnarray*}}

\def\gm5{\gamma^5}

\def\text#1{{\rm #1}}
\documentstyle[twoside,fleqn,espcrc2]{article}
\begin{document}
\title{Exactly massless fermions on the Lattice}
\author{Ting-Wai Chiu%
\address{Department of Physics, National Taiwan University,
Taipei, Taiwan 106, R.O.C.} \thanks{%
This work is supported by the National Science Council, R.O.C. under the
grant numbers NSC87-2112-M002-013 and NSC88-2112-M002-016}}

\begin{abstract}

The salient features of the Ginsparg-Wilson fermion in topologically
nontrivial background gauge fields are outlined.
The $R$-invariance of the zero modes, the indices and the index theorem
on a finite lattice are illustrated. The role of $ R $ in converting the
nonlocal $ D_c $ into a sequence of highly local $ D $ is demonstrated.

\end{abstract}

\maketitle

\section{INTRODUCTION}

My talk at Lattice 98 essentially consisted of two parts :
(1) The general solution of the Ginsparg-Wilson (GW) relation;
(2) Some numerical results for the ( generalized ) Neuberger-Dirac fermion
operator. In this writeup, due to the page limit, I confine my
discussions to the second part, which I did not have enough time to cover
in my talk. For the first part, I refer to ref. \cite{twc98:6a} and
ref. \cite{twc98:9a} for details.
Although the numerical results are obtained for the ( generalized )
Neuberger-Dirac fermion operator which is so far the only {\it explicit }
and {\it physical } solution of the GW relation, nevertheless,
they can serve the purpose to illustrate the general properties of
the GW fermion.

The salient feature of the GW fermion $ D $
\bea
D \gamma_5 + \gamma_5 D = 2 D R \gamma_5 D
\label{eq:gwr}
\eea
is that its chiral symmetry breaking part [ the RHS of (\ref{eq:gwr}) ]
does not change anything of the zero modes of $ D $ for any $ R $ which
controls the amount of chiral symmetry breaking as well as the locality of
$ D $. Therefore any zero mode of $ D $ must be a zero mode of $ D_c $,
the chiral symmetric fermion operator in the chiral limit $ R \to 0 $, and
vice versa. This can be easily seen by considering the general solution
\cite{twc98:6a,twc98:9a} of Eq. (\ref{eq:gwr})
\bea
D = D_c ( \Id + R D_c )^{-1} = ( \Id + D_c R )^{-1} D_c
\label{eq:gwf}
\eea
where
\bea
D_c \gamma_5 + \gamma_5 D_c = 0
\label{eq:Dc}
\eea
Consequently, $ N_{+}[D,A]=N_{+}[D_c,A] $ and $ N_{-}[D,A]=N_{-}[D_c,A] $
where $ N_{+}[D_c,A] $ ( $ N_{-}[D_c,A] $ ) is the
number of zero modes of positive ( negative ) chirality, a functional
of $ A $ and $ D_c $. The index of $ D $ is equal to the index of $ D_c $,
$ \mbox{ind}(D) \equiv N_{-}[D,A] - N_{+}[D,A]=\mbox{ind}(D_c) $.
These results hold for any $ R $.
According to the Nielson-Ninomiya no-go theorem \cite{no-go}, in the chiral
limit $ R \to 0 $, so if we want to keep $ D_c $ free of species doubling,
then $ D_c $ must be non-local. The non-locality is nothing to be afraid of,
and in fact we can live with it, and tame it to be harmless and even
useful. We will show that the zero modes of $ D_c $ agree with the continuum
solution excellently and the index theorem is realized exactly on a finite
lattice. Then we will turn on $ R $ and show that all topological
characteristics of $ D_c $ remain invariant while the nonlocal $ D_c $
is transformed into a sequence of local $ D's $. It has been shown in ref.
\cite{twc98:9a} that any well defined $ D_c $ must have zero index. In order
to have nonzero indices, $ D_c $ must be {\it singular ( i.e., divergent ) }
in topologically nontrivial background gauge field.
Although all these properties have been discussed analytically
in ref. \cite{twc98:9a}, nevertheless, it is instructive to convey this
picture vividly through some numerical examples in this paper.
More details of these studies will be reported later \cite{twc98:10}.

\section{GENERALIZED NEUBERGER \\
OPERATOR}

Consider the Neuberger-Dirac operator \cite{hn97:7}
in the chiral limit ( $ R \to 0 $ )
\bea
D_c = 2 M \frac{ \Id + V }{ \Id - V }, \hspace{4mm}
V = D_w ( D_w^{\dagger} D_w )^{-1/2}
\label{eq:DcV}
\eea
where $ D_w $ is the Wilson-Dirac fermion operator with negative mass
$ -M $ and Wilson parameter $ r_w > 0 $
\bea
D_w = - M
      + \frac{1}{2} [ \gamma_{\mu} ( \nabstar{\mu} + \nab{\mu} ) -
                      r_w  \nabstar{\mu} \nab{\mu} ]
\eea
where $ \nab{\mu} $ and $ \nabstar{\mu} $ are the forward and
backward difference operators.
Using Eq. (\ref{eq:gwf}), we obtain the generalized Neuberger-Dirac operator
\bea
D = 2 M ( \Id + V ) [ ( \Id - V ) + 2 M R ( \Id + V ) ]^{-1}
\eea
For $ M \in ( 0, 2 r_w ) $, $ D $ is topologically non-trivial.
In the following, we restrict our discussions to the simplest $ R $
which is proportional to the identity in the position space and
trivial in the Dirac space.
It was first demonstrated in ref. \cite{twc98:4}
that for $ R = 1/2 $ and in two dimensional $ U(1) $ background gauge
fields, $ D $ reproduces exact zero modes with definite chirality
on a finite lattice, the index theorem is satisfied exactly,
and the fermionic determinants are also in good agreement with the
continuum exact solutions.
In this paper, we again fix $ M = 1 $ and $ r_w = 1 $, and
explicitly verify that the Atiyah-Singer index theorem
\beq
Q = N_{-}[D,A] - N_{+}[D,A]
\label{eq:index_thm}
\eeq
holds exactly for all values of $ R $, say, from $ 10^{-6} $ to $ 10^{6} $,
in topologically nontrivial $ U(1) $ and $ SU(2) $ background gauge fields,
on two dimensional and four dimensional lattices respectively.
The two dimensional $ U(1) $ background gauge field has been discussed in
ref. \cite{twc98:4} and here we use the same notations as defined in
Eqs. (7)-(11) in ref. \cite{twc98:4}. We refer to ref. \cite{twc98:10}
for details of four dimensional results.

\section{NUMERICAL EXAMPLES}

In Fig. \ref{fig:zmode}, the exact zero mode with negative chirality
for $ D $ in topological charge $ Q = 1 $ background gauge field on a
$ 12 \times 12 $ lattice with periodic boundary conditions is plotted.
The agreement with the continuum exact solution \cite{sachs_wipf} is
excellent for any value of $ R $ ranging from $ 10^{-6} $ to $ 10^{6} $.
This verifies that the zero modes of any Ginsparg-Wilson fermion $ D $
are $ R $-invariant. It is striking to see that the zero mode of
$ D $ at $ R = 10^{-6} $, so close to the chiral limit, is still in
excellent agreement with the continuum exact solution.
In fact, the nonlocality of $ D_c $ does not prevent it
from having exact zero modes and satisfying the index theorem.
Since at the chiral limit, the Ginsparg-Wilson
relation is turned off, the existence of such a topologically nontrivial
$ D_c $ of course cannot be answered by the GW relation. However, if
we have already found such a $ D_c $, then we can tame its nonlocality
and singularity using Eq. (\ref{eq:gwf}) by choosing a sufficiently large
$ R $. In Fig. \ref{fig:local}, the absolute value of matrix elements,
$ | D(x,y) | $, versus $ | x - y | $ are plotted for a topologically
nontrivial gauge configuration on a $ 8 \times 8 $ lattice with
antiperiodic boundary conditions. At $ R = 10^{-6} $, $ D $ is nonlocal
and some elements become very large. Then at $ R = 0.1 $, $ D $ has been
tamed a lot but it is still far from local. For $ R > 0.5 $, $ D $
has become local and can be fitted by an exponential decay function.
If we increase $ R $ further, say, $ R=10.0 $, then $ D $ becomes almost
ultralocal, i.e., $ D(x,x+\hat\mu) \sim 10^{-3} \times D(x,x) $,
$ D(x,x+\hat\mu+\hat\nu) \sim 10^{-3} \times D(x,x) $, and other matrix
elements $ \sim 10^{-6} \times D(x,x) $. After $ D $ becomes local, the
fermion propagator remains almost the same for any larger $ R's $.
In Fig. \ref{fig:sl}, we plot the ( subtracted ) fermion propagator
$ D^{-1} (x,y) = \gamma_\mu S_\mu (x,y) $
( cf. Eq. (20) of ref. \cite{twc98:4} ) for $ R = 10.0 $
on a $ 8 \times 8 $ lattice, in a topologically trivial
background gauge field with $ h_1 = 0.1 $ and $ h_2 = 0.2 $.
The agreement with the continuum exact solution ( solid lines ) is excellent.

\psfigure 3.0in -0.3in {fig:zmode} {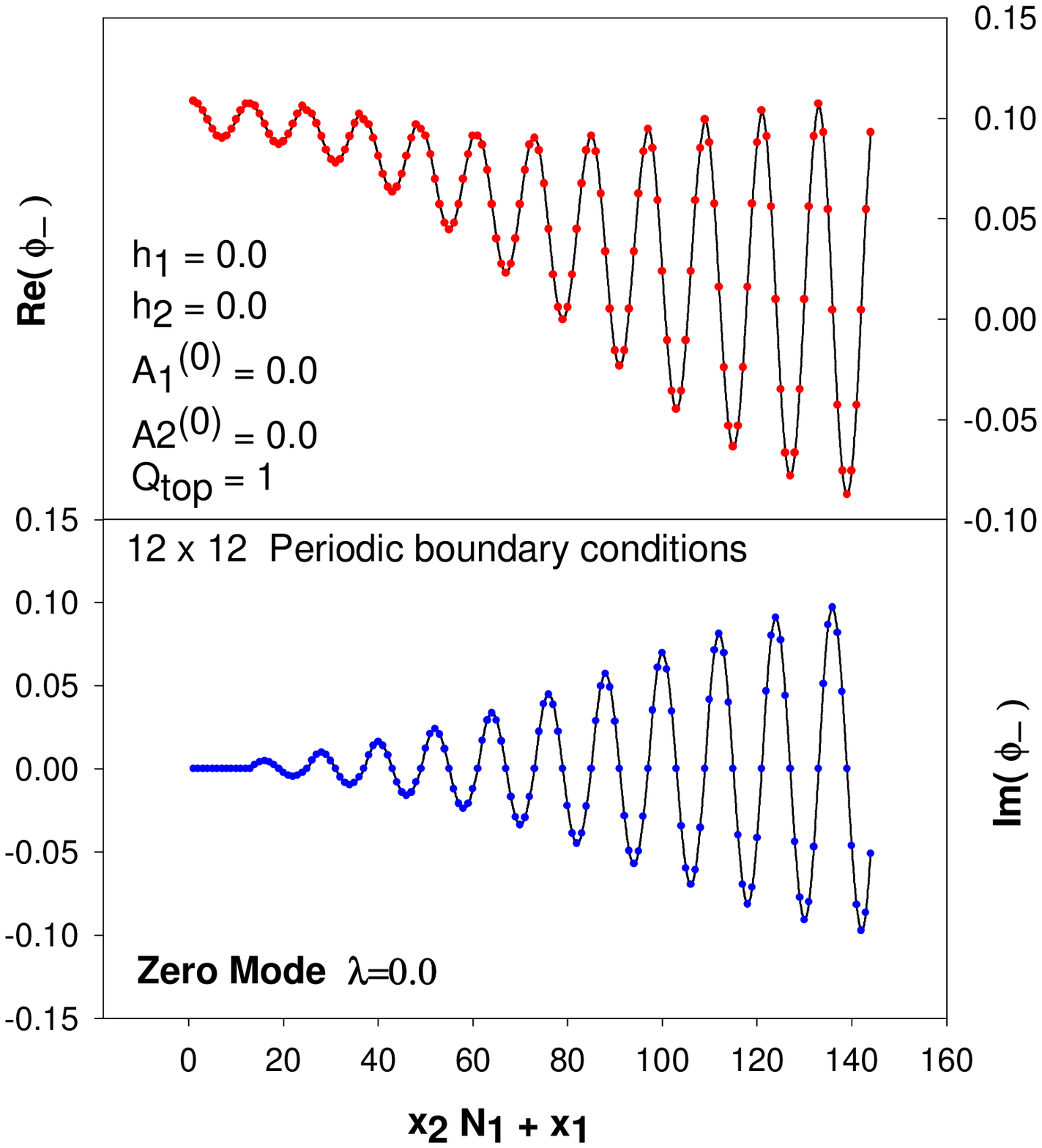} {
The real part and imaginary part of the zero mode of $ D $ for $ R=10^{-6} $.
They remain the same for all values of $ R $ from $ 10^{-6} $ to $ 10^6 $.
The solid lines denote the continuum exact solution.
}

\psfigure 3.0in -0.4in {fig:sl} {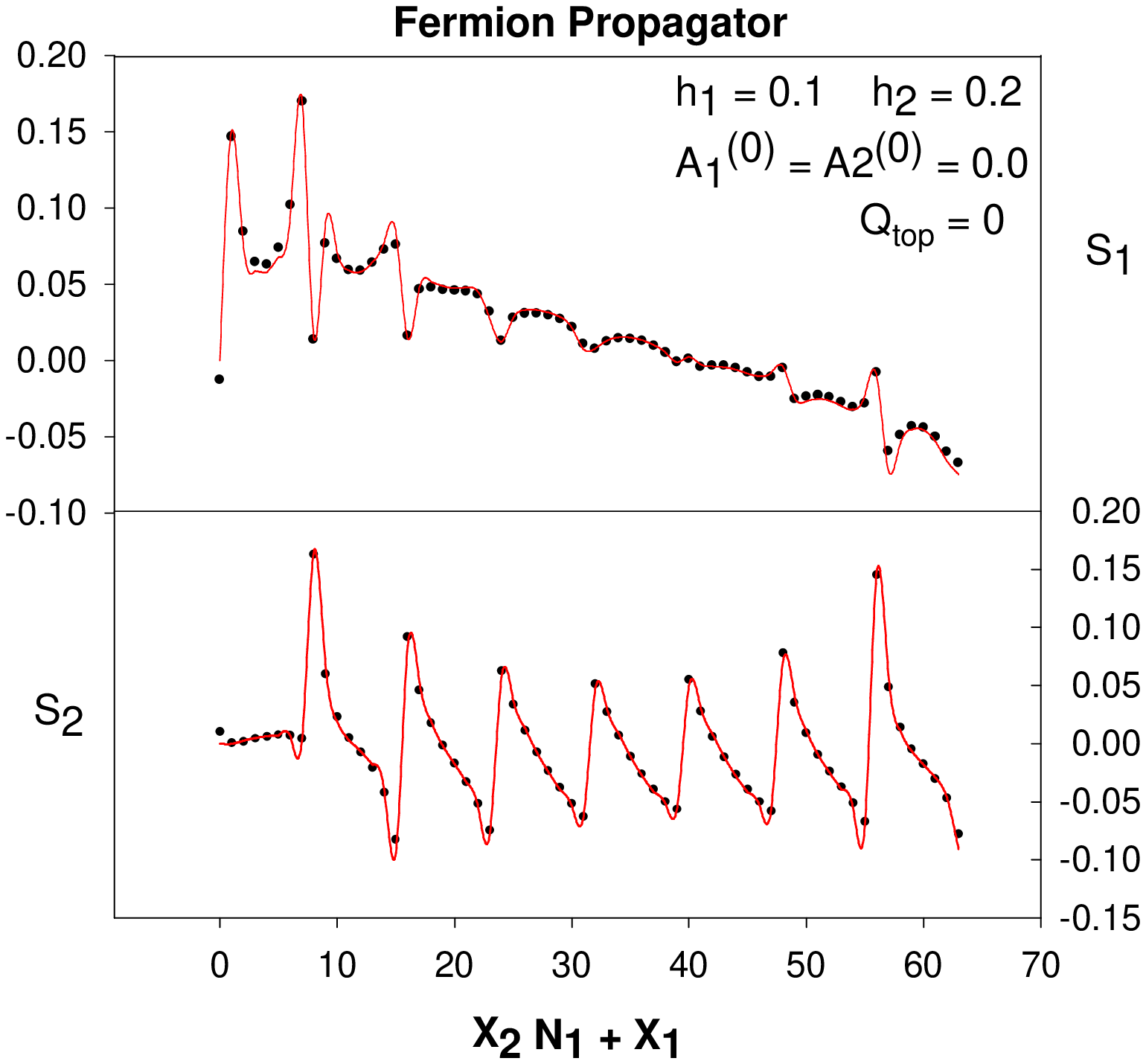} {
The fermion propagator $ D^{-1} (x,0) = \gamma_\mu S_\mu(x,0) $
for $ R = 10.0 $ on a $ 8 \times 8 $ lattice with antiperiodic
boundary conditions.
}

\psfigure 3.0in -0.3in {fig:local} {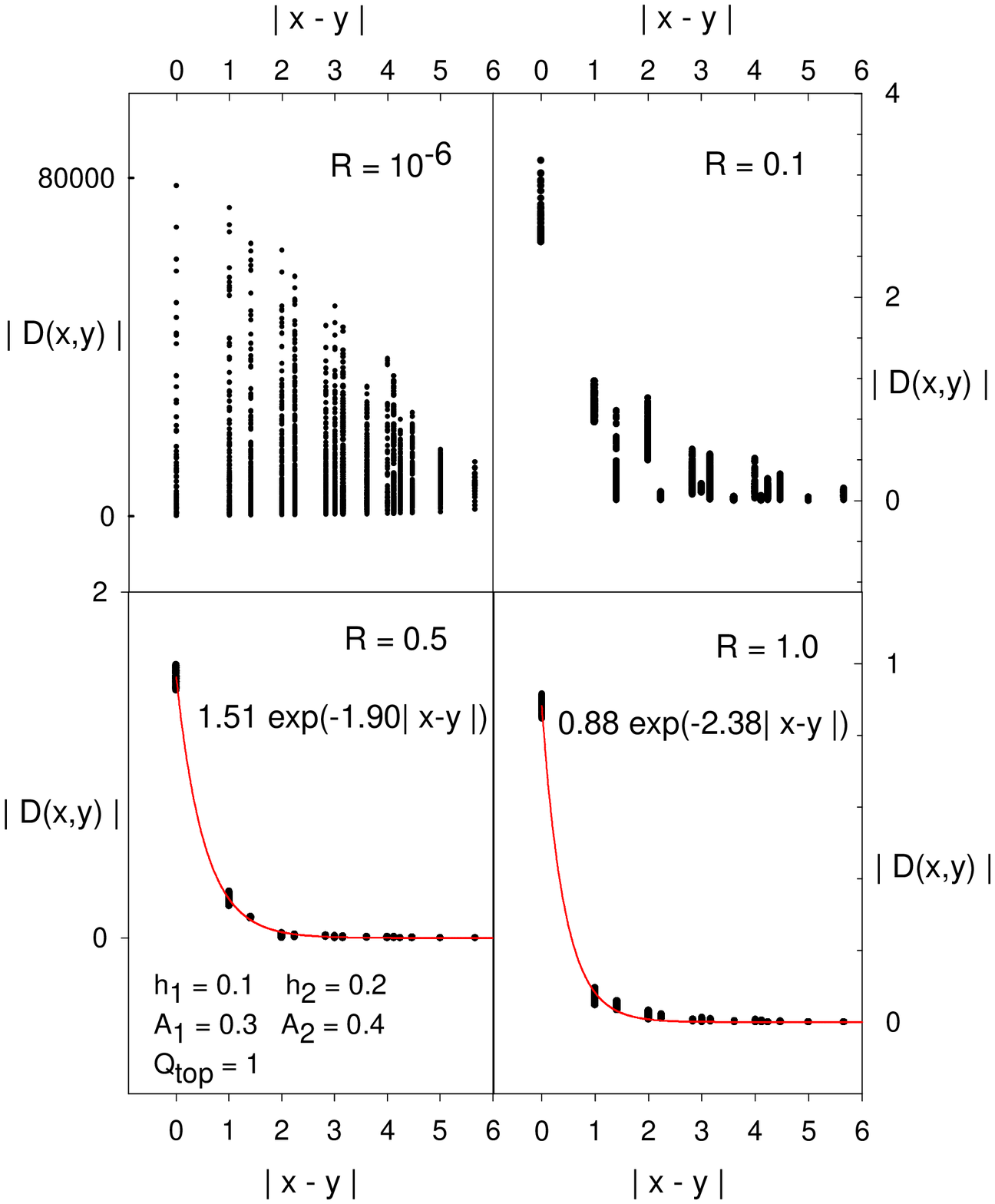} {
The locality of $ D $ versus $ R $ in a topologically nontrivial background
gauge field.
}

\section{SUMMARY and CONCLUSIONS}

To summarize, we have demonstrated that the nonlocality of a chirally
symmetric Dirac operator $ D_c $ on the lattice is one of the necessary
requirements for it to be free of species doubling and topologically
nontrivial, i.e., to have exact zero modes and
to realize the Atiyah-Singer index theorem on the lattice. The best way to
tame the nonlocality of $ D_c $ is to break the chiral symmetry via
the Ginsparg-Wilson solution Eq. (\ref{eq:gwf}) which gaurantees that
the topological characteristics of $ D_c $ is invariant for
any $ R $. In principle, we can make $ D $ almost ultralocal by simply
increasing $ R $.
Once $ D $ becomes local enough,
the fermion propagator $ D^{-1} (x,y) $ remains almost the same for all larger
$ R's $ except the diagonal elements
$ D^{-1}_{\alpha\alpha} (x,x) = R $, which can be easily
subtracted off. The $ R $-invariance of the zero modes and the Atiyah-Singer
index theorem has been verified for any value of $ R $ ranging from
$ 10^{-6} $ to $ 10^6 $.

\end{document}